# Femtosecond resolved mega-gauss magnetic field evolution in an intense laser generated solid plasma

Arvinder Singh Sandhu, Aditya Dharmadhikari, Paramel Pattathil Rajeev and G. Ravindra Kumar
Tata Institute of Fundamental Research, Dr. Homi Bhabha Road, Colaba, Mumbai 400 005.

**Fax: =91-22-2152110; E-mail: grk@tifr.res.in**

## ABSTRACT

We report the first femtosecond resolved evolution of the giant magnetic field in a solid plasma, produced by a 100 fs, $10^{16}$ W cm$^{-2}$, 806 nm laser field, using a pump-probe Faraday rotation technique.

PACS Numbers:

One of the important features of the interaction of an intense laser with a solid is the creation of giant magnetic fields in the ensuing plasma[1]. There is a great deal of interest in the occurrence of huge quasi-steady state ("DC") magnetic fields in such plasmas due to the potential of hybrid (inertial and magnetic) confinement in laser fusion[2] and because of the fast igniter proposal[3]. Giant magnetic fields play a crucial role in the transport of electrons, in particular the hot electrons that arise at high intensities. A recent report on isotope separation from solid plasmas[4] cites magnetic fields as a possible source of ion separation `centrifuge', though the actual mechanisms may be more complicated[5]. Since the first observation of such magnetic fields by Stamper et al.[6], their origin, magnitudes and qualitative features have attracted considerable attention and experiments have used different techniques like Faraday rotation[6], Zeeman profiling[7], current probes[8], magnetic induction in a coil[9] and demagnetization of a predefined magnetic domain [10]. Most of the experiments till date have used nanosecond lasers to map out the field distributions and significant progress has been made in the experimental study of toroidal, axial and poloidal field generation. The magnitude and direction of these fields obviously depend on the plasma generation mechanisms, which in turn depend crucially on the polarization and angle of incidence of the incident laser beam. Many mechanisms – Nernst (thermoelectric) effect[6], ponderomotive acceleration[11], dynamo effect etc. have been examined as the sources of these magnetic fields[1].

A brief summary of some of the previous experimental and theoretical work is as follows. Stamper et al.[6] used Faraday rotation in a number of experiments to map out the magnetic fields. Kieffer and coworkers[12] measured axial magnetic fields of 0.6 MG based on the dynamo effect at normal incidence of a 100 ps, $10^{15}$ W cm$^{-2}$ pulse on an aluminium target. Luther-Davies and coworkers [13] mapped out the spatial and temporal evolution for 20 ps pulses magnetic field using interferometry and polarimetry and concluded that the maximum value was attained within 0 to 30 ps of the peak of the laser pulse. They found that the field reduced below their detection limit of 250 kG in 100 ps. Haines[14] has discussed the role of magnetic fields arising from different mechanisms in hot electron transport and its impact on laser fusion. Eliezer and coworkers[15] measured Faraday rotation from plasma produced at normal incidence by a circularly polarized laser field. Khan et al.[16] have discussed the role of ponderomotive action in the generation of axial magnetic fields measured using rotation of the stimulated Brillouin scattering (SBS) signal from the plasma.

On the theoretical front, Nishihara et al[17] calculated the magnitudes of magnetic field and its time evolution for resonance absorption. Bezzerides et al[18], Speziale and Catto[19] and Woo and DeGroot[20] later considered more extensively, resonance absorption and other mechanisms of magnetic field generation. Chakaraborty et al[21]. have reported scaling laws for axial magnetic field in terms of the laser wavelength, power and electron density. Recent theoretical work includes the consideration of resonance absorption at relativistic intensities by Mima and coworkers[22] and an analytical model for the generation of poloidal and toroidal field generation by Bhattacharyya[23] and coworkers.

Recently, picosecond and femtosecond laser produced solid plasmas have provided a new experimental situation for the study of these magnetic fields. It is now well established that femtosecond laser produced plasmas are qualitatively different from those generated by nanosecond laser fields[24] - hydrodynamic expansion is insignificant during the interaction leading to higher electron densities and temperatures. It is expected that magnetic fields should also be qualitatively different in such plasmas. Apart from collisional and resonance absorption mechanisms, vacuum heating is also likely to be a mechanism of plasma excitation in the femtosecond regime. More interestingly, an the attractive possibility arises – that of creating a giant magnetic field at higher intensities and monitoring the evolution of that field on ultrashort time scales. Borghesi et al have [25] recently reported, albeit in a broad fashion, the first study of the time evolution of such magnetic fields on picosecond time scales, using a laser of 1.5 ps duration at normal incidence on a target. We present here, the detailed temporal evolution of these megagauss magnetic fields at femtosecond resolution and attempt an interpretation of the observations. As will be discussed later our setup defines zero delay with better precision, thus enabling us to look at initial buildup of Magnetic field.

We use the technique of Faraday rotation in our experiment. The advantage of this technique is that it causes the least disturbance to the plasma and serves as a remote diagnostic, besides possessing the inherent high sensitivity of an optical technique. Unlike all previous experiments, we use obliquely incident pump and probe. Experimentally, this provides the additional advantage of getting the probe to penetrate up to the critical layer of the plasma. Further, in this geometry, mechanisms like resonance absorption[26] and vacuum heating[27] are expected to play a very important role in addition to collisional absorption. We monitor the polarization of a weak probe beam that interacts with the plasma created by a strong pump pulse. The changes in probe polarization reflect the evolution of the magnetic field created within the plasma. For this technique to work effectively, the amount of diffuse scatter of the laser beams should be minimal, as is indeed the case for femtosecond pulse created plasmas which offer a nearly planar geometry. We observe megagauss fields whose magnitude depends on the incident laser intensity as well as the target material. We also report preliminary measurements of pump polarization dependence of magnetic field generation.

Figure 1 shows a sketch of our experimental set up. We obtain a high intensity ($>10^{16}$ W/cm$^2$) 100fs pump pulse from a custom built chirped pulse amplification Ti:S laser which is described in detail elsewhere[28]. The photodiode in the input beam PD3 serves to measure the fluctuations in input laser beam. Then beam splitter BS1 splits laser into two parts, the linearly polarized pump pulse is incident at 50 degrees on the metal targets to generate the plasma, and a linearly polarized probe pulse incident at 55 degrees, which is made hundred fold weaker compared to pump. The plasma is generated by pump pulse within 100 fs and has a very steep density gradient initially. This is established by monitoring the Doppler shift of the wavelength of reflected pump light from which we infer the hydrodynamic expansion velocity (See Fig. 3). We monitor the amplitude and polarization of the specularly reflected probe. We observed that there was no diffuse scatter of the probe or pump within our range of observations. The targets are housed in a chamber evacuated to $10^{-3}$ torr and are mounted on a rotation and vertical translation stage. Thus after every shot a fresh surface of the metal is exposed to laser. The targets used in these experiments were polished Cu and Al discs with flatness better than λ/4. The change of probe polarization by Faraday rotation due to the magnetic field present in the plasma is studied as a function of the time delay between pump and probe. The pump-probe delay time has resolution of 1 micron, the probe path having a motorized translation stage. We have a 10fs resolution of the delay between pump and probe and can monitor the temporal behavior of polarization rotation from zero time to 50 ps time delay. Using zero order thin half wave plates we can change the polarization of both pump and probe beams. We obtain best focusing of the laser beams by maximizing hard x-ray bremsstrahlung signals with a NaI (Tl) detector. The spots of both beams are spatially matched using a CCD camera with a 6X zoom. The technique used for getting temporal overlap involves monitoring of time resolved reflectivity. This apart from being a valuable experiment in itself also serves the purpose of normalization of reflected probe light to isolate polarization rotation signal from time dependent reflectivity change. The zero delay position is located by looking for the steep reflection dip as the probe starts reflecting from plasma generated by pump instead of metal target. This dip being very sharp (order 100fs) establishes zero very accurately. At negative delay, i.e. when the probe arrives earlier than pump, polarization of the probe beam remains unchanged after reflection. Spatial and temporal matching of the two beams was externally monitored during the experimental run, by splitting of input beams after passing through the lenses, onto a polyphenylene vinylene (PPV) film. The third order autocorrelation signal from PPV serves as a check on the spatial and temporal overlap in addition to characterizing the laser pulse.

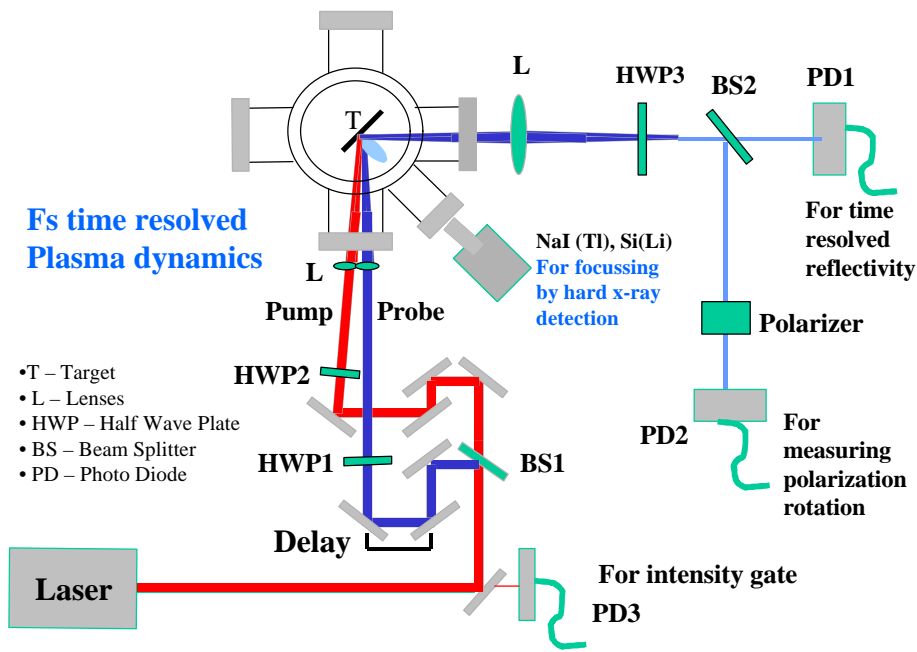

*Fig 1. Experimental set up*

The Faraday rotation diagnostic consists of lens to recollimate the reflected probe, a zero order half-wave plate (HWP3) to simulate Faraday rotation for calibration purposes, and a beam splitter, which divides the light almost equally into two separate arms having a similar photodiode each (PD1 and PD2). In one arm (PD2) we have put a polarizer and set it to null the signal (crossed wrt probe polarization state) and is fixed at that position. All three photodiode signals are simultaneously acquired for each laser shot along with each delay stage position in a computerized acquisition. Further the ratio of PD2/PD1 is also computed for position and this serves to measure magnitude of polarization rotation. Any rotation of polarization state will result in leaking of light through PD2, which was set to cross polarization state in absence of pump whereas signal in PD1 remains same. Thus the ratio PD2/PD1 will increase depending upon amount of rotation. The object of taking ratio is to cancel the change occurring due to time delay dependence of reflectivity and thus isolating only rotation behavior as a function of time delay. In order to convert ratio change to corresponding rotation in degrees we need to do the calibration. We did this by rotating the half wave plate HWP3 (in absence of pump) thus simulating the actual rotation of polarization and observing ratio change. We obtain the expected parabolic calibration curve of polarization rotation signal (ratio of PD2/PD1) wrt θ (artificial rotation). Then half wave plate is left at the position where the signal is crossed in the PD2. In an actual experimental run with pump on and the probe delay is continuously varied from –10ps to +40ps, reflectivity (PD1) and polarization rotation (PD2/PD1) as a function of time delay.

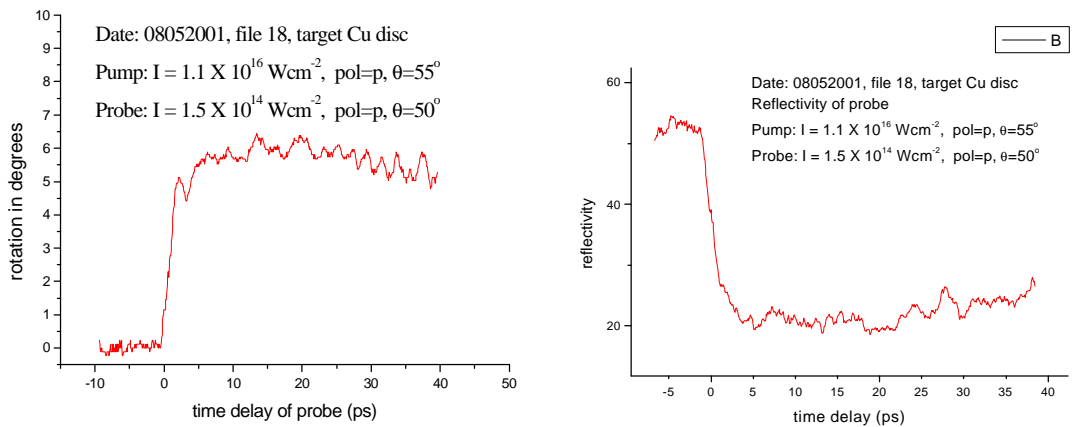

**Fig.2, 3: Variation of polarization rotation and reflectivity with respect to probe delay.**

We have thus taken time resolved polarization rotation data for various incident laser intensities and polarizations. In fig2 we show a typical raw data for polarization rotation in degrees, at pump intensity ($10^{16}$ Wcm$^{-2}$) for a copper target. Corresponding time resolved reflectivity is also shown in fig3. In order to check that increase of ratio (rotation signal) is real and has no contribution from optics in the path, we repeat the same run but without the polarizer in front of PD2. Fig4 shows the corresponding ratio behavior in that case, which is constant to a good degree. At high pump intensities the maximum rotation angle as seen from the fig2 data is about 6-7 degree. After initial increase the rotation then slowly decreases depending upon the intensity of the pump. This behavior was again double checked by actually measuring the ellipticity of probe (by rotating the polarizer) at different delays in steps of 5ps. We have carried out above measurements for different intensities in the range of $1 \times 10^{15}$ to $2 \times 10^{16}$ W cm$^{-2}$ for copper as well as aluminum targets. The maximum magnetic field as well as decay time of magnetic field (rotation signal) depends upon pump intensity. In case of aluminum we have done polarization rotation measurements for both s and p polarized pump.

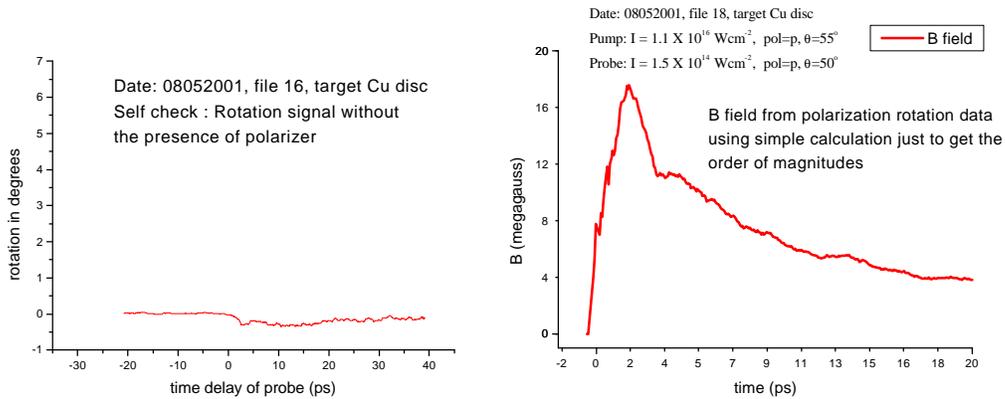

**Fig. 4, 5: Derived (PD2/PD1) `rotation signal' in absence of the analyzer; Estimated magnetic fields as a function of probe delay.**

Before discussing the results further we estimate the magnetic field from the Faraday rotation data in all the experiments. There is wide variety of quasi-static magnetic field generation mechanisms possible in any real experiment (as mentioned above) and it may be possible that more than one of these is important in our study. Without going into any specific mechanism, we attempt to get the dynamic evolution of "average magnetic" fields in the plasma. For the probe beam locally the amount of Faraday rotation $\theta \propto n_e B L$. However, for finite plasma the average rotation at a fixed time is given as,

$$\nabla q(t) = \frac{4p \bullet e^3}{2m^2 c^2 w^2} \int_0^{L(t)} n_e(l,t) B(l,t) dl ,$$

Where the $L(\tau)$ is the plasma length scale for given time delay the $\tau$.

Before going into detailed calculations we employ a simplistic picture to get approximate temporal evolution of magnetic field. Let us assume some average electron density inside plasma and average magnetic field, and then a simple estimate is given by

$$\nabla q(t) = 5.9 \times 10^{24} \bullet n_e(t) L(t) B(t).$$

Assuming L increases with time as plasma front expands with velocity $v_{exp} \approx 5 \times 10^6$ cm/s (estimated from Doppler shift measurements) and i.e. $L(\tau) = L(0) + v_{exp}\tau$. Then starting with $L(0) = 100$ nm and $<n_e> 10^{21}$/cc, we get magnetic fields as shown in fig5.